\documentclass[12pt]{article}
\usepackage{epsfig}
\usepackage{amssymb,graphicx}
\def\be{\begin{equation}}
\def\ee{\end{equation}}
\def\ba{\begin{array}{c}}
\def\ea{\end{array}}

\def\ben{\[}
\def\een{\]}

\newcommand{\bea}{\begin{eqnarray}}
\newcommand{\eea}{\end{eqnarray}}

\begin{document}

\begin{center}

{\Large \bf {

Symmetrized exponential oscillator

 }}

\vspace{13mm}

 {\bf Miloslav Znojil}

 \vspace{3mm}
Nuclear Physics Institute of the CAS, Hlavn\'{\i} 130, 250 68 \v{R}e\v{z},
Czech Republic

\vspace{3mm} and

 \vspace{3mm}

 Institute of Systems Science, Durban University of
Technology, 4000 Durban, South Africa

\vspace{3mm}

 e-mail:
  znojil@ujf.cas.cz

\vspace{3mm}


\end{center}

\subsection*{Keywords:}

.

quantum bound states;

exactly solvable models;

Bessel special functions;

transcendental secular equation;

numerical precision;

\subsection*{PACS numbers:}
.

PACS 03.65.Ge – Solutions of wave equations: bound states

PACS 02.30.Gp – Special functions



\section*{Abstract}

Several properties of bound states in potential $ V(x)= g^2\exp
(|x|)$ are studied. Firstly, with the emphasis on the reliability of
our arbitrary-precision construction, wave functions are considered
in the two alternative (viz., asymptotically decreasing or regular)
exact Bessel-function forms which obey the asymptotic or matching
conditions, respectively. The merits of the resulting complementary
transcendental secular equation approaches are compared and their
applicability is discussed.

\newpage

\section{Introduction\label{I}}

\begin{figure}[h]                    
\begin{center}                         
\epsfig{file=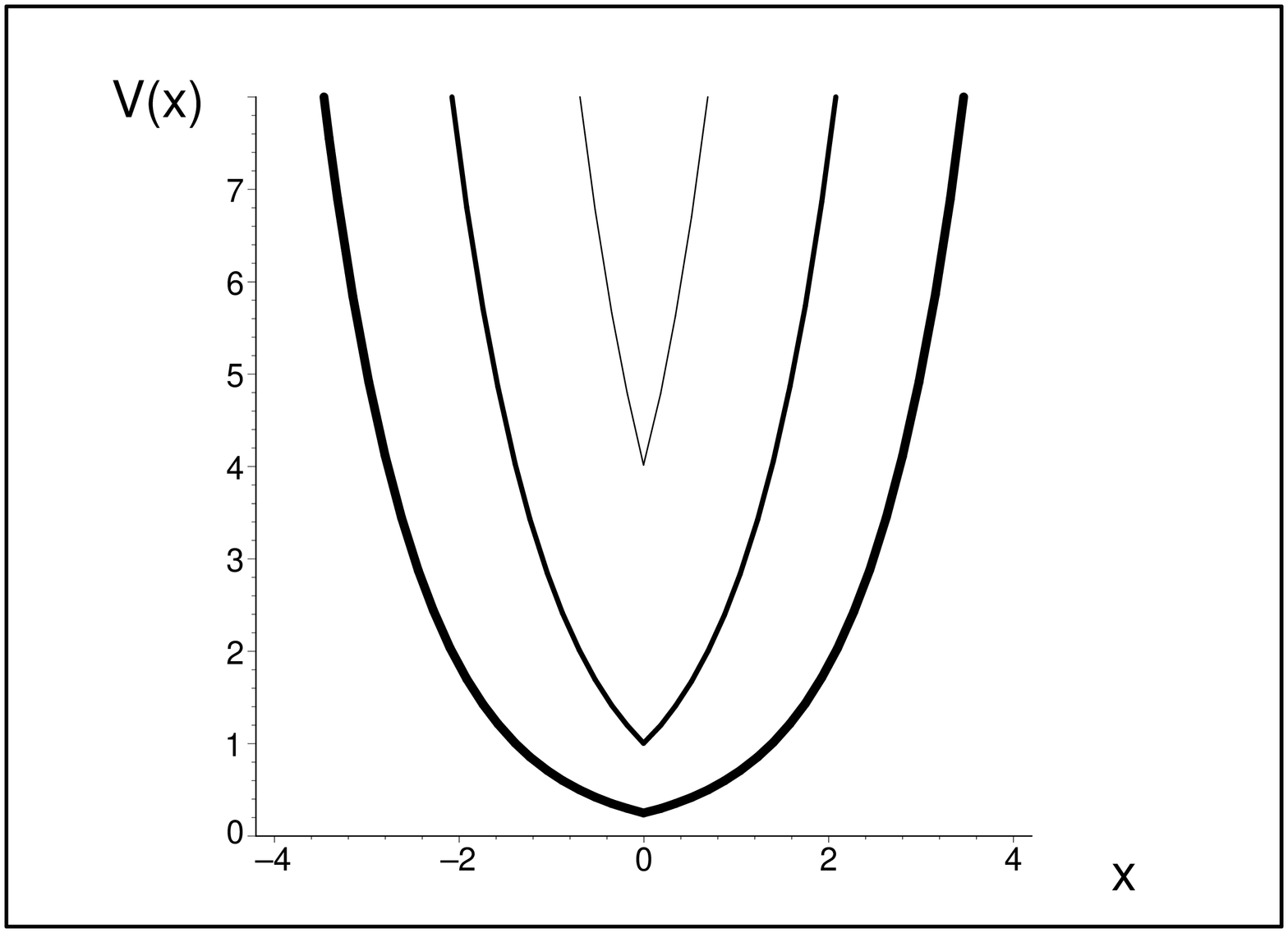,angle=270,width=0.36\textwidth}
\end{center}    
\vspace{2mm} \caption{The shape of potential (\ref{exwe}) at small
$g^2=1/4$ (thick, almost smooth-looking curve), medium $g^2=1$
(medium curve) and larger $g^2=4$ (thin curve, sharply spiked
shape).
 \label{reqizt}
 }
\end{figure}

One-dimensional bound-state Schr\"{o}dinger equation
 \be
 -\, \frac{{\rm d}^2}{{\rm d} x^2} \psi_n(x)
 + V(x) \psi_n(x)= E_n\,
 \psi_n(x)\,,
 \ \ \ \
 \psi_n(x) \in L^2(\mathbb{R})\,,
 \ \ \ \ n = 0, 1, \ldots
 \,
   \label{SEx}
  \ee
with the centrally symmetric confining potential
 \be
 V(x)= g^2\exp (|x|)\,,\ \ \ \ \ x \in (-\infty,\infty)
 \label{exwe}
 \ee
(cf. Fig.~\ref{reqizt}) does not seem to have attracted attention of
the authors of textbooks on quantum mechanics. In the context of
quantum phenomenology, such a neglect is certainly undeserved since
the asymptotic growth of the potential may be perceived as
positioned somewhere in between quadratic harmonic-oscillator
$V^{(HO)}(x)= (\omega\,x)^2$ with equidistant, vibration-type
spectrum of energies $E_n = (2n+1)\omega $, $n = 0, 1, \ldots$ and
its equally popular infinite-power square-well partner $V^{(SW)}(x)=
(\omega\,x)^\infty$ leading to the quadratic growth of $E_n = \pi^2
(n+1)^2\omega^2 /4$, $n = 0, 1, \ldots$ resembling the rotational
energy bands \cite{Constantinescu}.

The neglect of model (\ref{exwe}) seems equally undeserved from the
purely formal point of view because the related Schr\"{o}dinger
equation is extremely elementary and exactly solvable in terms of
the modified Bessel functions matched in the origin
\cite{Wolf,rese}. The low popularity of the conceptual as well as
practical use of the bound-state families as described by
Eqs.~(\ref{SEx}) and (\ref{exwe}) may be given several tentative
explanations. We shall return to this point later, in section
\ref{discussion} below. Now, let us only mention the feature of
non-analyticity of the potential in the origin. Indeed, the
analyticity along the whole real line is precisely what makes
harmonic oscillators so popular, say, as toy models in quantum field
theory \cite{fiethe} as well as in rigorous functional analysis
\cite{FHS,FHSb}.

In what follows we intend to re-attract attention to the
non-analytic model (\ref{SEx}) + (\ref{exwe}) and to describe and
discuss a few basic properties of its exact solutions at some
length, therefore. Let us also add that in a historical perspective
our present interest in a deeper study of the ``missing link''
(\ref{exwe}) connecting $V^{(HO)}(x)$ with $V^{(SW)}(x)$ found its
origin in the methodical aspects of quantum gravity and, in
particular, in the related Smilga's comment \cite{Smilga} on the
dynamics of the so called Pais-Uhlenbeck toy-model oscillator.
Indeed, in the latter model the dynamics is mimicked using a pair of
potentials $V^{(HO)}(x)$ with different $\omega$s (cf. Eqs. Nr. 5
and 6 in {\it loc. cit.}). In our most recent paper \cite{Markov} we
tried to test some of the Smilga's conclusions using an alternative,
Pais-Uhlenbeck-like model in which the dynamics was generated by
square-wells  $V^{(SW)}(x)$ with different $\omega$s (cf. Eqs. Nr. 1
and 2 in {\it loc. cit.}). Unfortunately, the results of the
comparison proved too model-dependent. In other words, the
square-well simulation of the Pais-Uhlenbeck-related effects were,
from the point of view of the original problem of quantum gravity,
next to useless. Thus, the study of the present model (\ref{exwe})
can be perceived as offering a new hope and having a rather deep
motivation, especially due to its exact solvability.

\section{Mathematical aspects of the model\label{II}}

\subsection{Exact solvability}

On the Wolfram's webpage \cite{Wolf} one easily finds that
differential Eq.~(\ref{SEx}) + (\ref{exwe}) re-written in the form
 \be
 -y''(x)+a\, \exp (b x)\, y(x) = c\, y(x)
 \label{simplif}
 \ee
has general solution
 $$
 y(x) = k_1 e^{\pi \sqrt{c}/b} \Gamma(1-2 i \sqrt{c}/b) I_{-2 i
 \sqrt{c}/b}\left (2 \sqrt{a\, e^{b x}}/b \right )+
 $$
 $$
 +k_2 e^{-\pi \sqrt{c}/b}
 \Gamma(1+2 i \sqrt{c}/b)  I_{2 i
 \sqrt{c}/b}\left (2 \sqrt{a\, e^{b x}}/b \right )
 $$
where $I_n(z)$ denotes the modified Bessel function of the first
kind and where $\Gamma(x)$ is the gamma function.

It is well (and for a long time,  \cite{Ma}) known that the
existence of the latter closed-form solutions renders the
construction of the states which are generated by the present
confining interaction (\ref{exwe}) non-numerical (or semi-numerical
at worst, cf. \cite{rese}). What is much less often acknowledged is
that the special-function solvability of the problem opens also the
way towards various forms of its innovative deformations. Let us
recall here just two illustrative examples. In the first one one can
require a complete survival of the quantum stability of the system
in question (i.e., of the strict reality of the energy spectrum)
even after the real potential gets replaced by its suitable
complexified form. In Ref.~\cite{CJT}, for example, the authors
achieved such a goal via an elementary replacement $b \to {\rm
i}\,b$ in Eq.~(\ref{simplif}) (cf. {\it loc. cit.\,} for more
details).

In the role of the second illustrative example of the change of
horizons based on the analytic continuation techniques we may recall
the change of parameters $(a,b) \to \ (-a,-b)$ in
Eq.~(\ref{simplif}) which is of interest in molecular physics. In
such an alternative dynamical regime one studies, typically, the
resonances (i.e., complex eigenvalues) which are generated by a
``realistic'', asymptotically vanishing real potential with a
repulsive core in the origin (cf., e.g., Ref.~\cite{Marcelo} and
references therein).

In what follows our attention will remain restricted to the most
straightforward bound-state interpretation (\ref{SEx}) +
(\ref{exwe}) of the exponential interaction and of the Bessel's form
of differential Eq.~(\ref{simplif}).

\subsection{Consistency of the non-analyticity at $x=0$}

In the spirit of our recent brief comment \cite{star} on ordinary
differential Schr\"{o}dinger equations let us first remind the
readers that the existence of the single and isolated point of
non-analyticity of any given potential $V(x)$ in the origin may be
given a very natural interpretation in the context of the theory of
quantum graphs. In this language, any one-dimensional
Schr\"{o}dinger equation may be identified with a quantum graph with
one vertex and two edges. In practice this means that the equation
is split into a pair of half-line differential equations
 \be
 -\, \frac{{\rm d}^2}{{\rm d} s^2} \psi_{(left)}(s,E)
 +  g^2 e^{-s}\,\psi_{(left)}(s,E)= E\,\psi_{(left)}(s,E)\,,
 \ \ \ \ s \in (-\infty,0)\,,
   \label{SExexleft}
  \ee
 \be
 -\, \frac{{\rm d}^2}{{\rm d} r^2} \psi_{(right)}(r,E)
 +  g^2 e^r\,\psi_{(right)}(r,E)= E\,\psi_{(right)}(r,E)\,,
 \ \ \ \ r \in (0,\infty)
   \label{SExexright}
  \ee
such that the logarithmic derivatives of the two respective halves
of the wave function are properly matched in the origin,
 \be
 \psi_{(left)}(0,E)=
 \psi_{(right)}(0,E)\,,
 \ \ \ \ \
 \psi'_{(left)}(0,E)=
 \psi'_{(right)}(0,E)\,.
 \ee
Such a split of $\psi_n({x})$ into two halves reflects the spatial
symmetry of the interaction, $V(x)= V(-x)$ so that we are allowed to
restrict attention, say, just to the right half-axis (so we may drop
the subscript ``right'' as redundant), distinguishing just between
the conventionally normalized even-parity bound states such that
 \be
 \psi_{(even)}(0,E)=1\,,
 \ \ \ \ \
 \psi'_{(even)}(0,E)=0\,,
 \ \ \ \ E=E_{n}\,,
 \ \ \ n=0,2,4,\ldots
 \label{evensec}
 \ee
and the odd-parity bound states such that
 \be
 \psi_{(odd)}(0,E)=0\,,
 \ \ \ \ \
 \psi'_{(odd)}(0,E)=1\,,
 \ \ \ \ E=E_{n}\,,
 \ \ \ n=1,3,5,\ldots\,.
 \label{oddsec}
 \ee
Let us add that the left-right split of the problem may be also
perceived as a one-dimensional parallel to the partial-wave
expansion of the wave functions in three dimensions. In such a
perspective one of Eqs.~(\ref{SExexleft}) or (\ref{SExexright})
plays the role of an analogue of the radial Schr\"{o}dinger
equation. Similarly, the operator of parity ${\cal P}$ may be
perceived as a one-dimensional counterpart of the rotational
symmetry of the three-dimensional central $V(\vec{x}) =
V(|\vec{x}|)$.

In the same spirit, the three-dimensional quantum number of angular
momentum $\ell = 0, 1, \ldots$ degenerates here to the two values of
the parity quantum number $\pm 1$. Thus, we may conclude that the
non-analyticity of our present potential (\ref{exwe}) in the origin
finds its analogue in the optional central singularities exhibited,
e.g., by the Coulombic attraction $V(|\vec{x}|) \sim -1/|\vec{x}|$
or by the more singular repulsive cores (e.g., $V(|\vec{x}|) \sim
|\vec{x}|^{-6}$ \cite{singular}) in three dimensions.

\subsection{Asymptotically decreasing solutions and the exact energies}

Via the change of independent variable $r \to y = \exp (r/2)$ one
transforms the radial-like differential Eq.~(\ref{SExexright})
(living, at positive energies $E>0$, on the half-line of coordinates
$r = |x| \in (0,\infty)$) into another differential equation which
is solvable in terms of Hankel functions ({\it alias} Bessel
functions of the third kind -- see their definition in \cite{nist}).

At an arbitrary trial energy $E=k^2$ this yields all of the
admissible general solutions of Eq.~(\ref{SExexright}) in the
compact, exact and explicit special-function form
 \be
 \psi_{(general)}(r,k^2)=
 C_1\,H^{(1)}_{\nu}(z)+
 C_2\,H^{(2)}_{\nu}(z)\,,
 \ \ \ \ z = 2ige^{r/2}\,,
 \ \ \ \ \nu = 2ik\,.
 \label{haha}
 \ee
Once we require, in the bound-state context, the asympotically
decreasing behavior of these solutions we may consult formulae
10.2.5 and 10.2.6 in Ref.~\cite{nist} and conclude that we must put
$C_2=0$. The resulting, asymptotically correct solutions have the
compact and final first-Hankel-function form
 \be
 \psi_{(asymptotically\ decreasing)}(r,k^2) \sim
  H^{(1)}_{2ik}\left (2ige^{r/2} \right )\,.
  \label{hlejost}
  \ee
Naturally, this solution varies with the trial-energy variable
$E=k^2$ but the value of this variable must be fixed via the
transcendental secular equations (\ref{evensec}) (even-parity
bound-state energies) or (\ref{oddsec}) (odd-parity bound-state
energies), i.e., via the respective transcendental-equation
constraint at $r=0$, viz.,
 \be
 \frac{d}{dg}\,
 H^{(1)}_{2ik}\left (2ig \right )=
 0\,,\ \ \
 \ \ \ n = 0, 2, 4, \ldots
 \label{evensol}
 \ee
or
 \be
 H^{(1)}_{2ik}\left (2ig \right )=0\,,\ \ \
 \ \ \ n =1, 3, 5, \ldots\,.
 \label{oddsol}
 \ee
These equations must be solved numerically of course. The former,
slightly more complicated one may be still given the two alternative
but equivalent simplified forms, viz.,
 \be
 H^{(1)}_{2ik-1}\left (2ig \right )-\frac{g}{k}\,H^{(1)}_{2ik}\left
(2ig \right )
 =0\,,\ \ \
 \ \ \ n = 0, 2, 4, \ldots
 \label{evensolb}
 \ee
or
 \be
 H^{(1)}_{2ik-1}\left (2ig \right )-H^{(1)}_{2ik+1}\left (2ig \right)
 =0\,,\ \ \
 \ \ \ n = 0, 2, 4, \ldots\,.
 \label{evensolc}
 \ee
Both of these simplifications follow from formula 10. 6. 1 in
\cite{nist}. The construction is complete.

\section{Physical aspects of the model\label{III}}

Although the detailed and exhaustive analysis of the precision of
the numerical determination of the bound-state energies $E=k^2$ via
secular equations (\ref{evensol}) - (\ref{evensolc}) lies far beyond
the scope of the present letter, several features of these equations
deserve an explicit commentary.

\begin{figure}[h]                    
\begin{center}                         
\epsfig{file=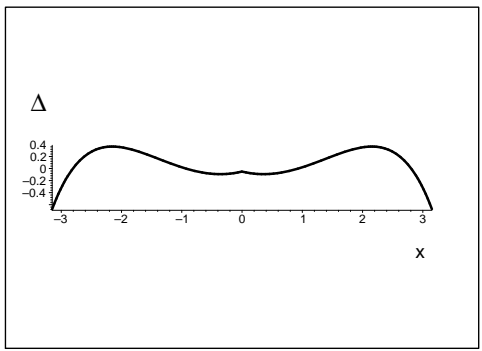,angle=270,width=0.36\textwidth}
\end{center}    
\vspace{2mm} \caption{The short-range smallness of difference
between harmonic oscillator and exponential potential (\ref{exwe})
($\Delta=0.5\cdot x^2+0.2-g^2\cdot \exp |x|\ $ with $\ g^2=1/4$).
 \label{aqizt}
 }
\end{figure}

\subsection{The loss of precision in the domain of very small couplings $g$ }

From our illustrative Fig.~\ref{reqizt} (and, in particular, from
its complement Fig.~\ref{aqizt}) one could deduce that in the
dynamical regime of very small couplings $g$ the shape of our
potential (\ref{exwe}) is not too different from the shape of
harmonic oscillator, within the range of the low-lying spectrum at
least. Hence, the exponential-well spectrum could also, in some
sense, resemble the spectrum of the harmonic oscillator. Naturally,
this does not imply that its search must necessarily be also
numerically well behaved. A word of warning may be found, e.g., in
Ref.~\cite{Marcelo} in which it has been argued that in the
numerical practice and, in particular, in the case of exponential
potentials it is often useful to test the precision of the numerical
root-searching results using an alternative algorithm (in
particular, in {\it loc. cit.\,} the authors recommended the use of
the so called Riccati-Pad\'{e} method).

%
%
\begin{figure}[h]                    
\begin{center}                         
\epsfig{file=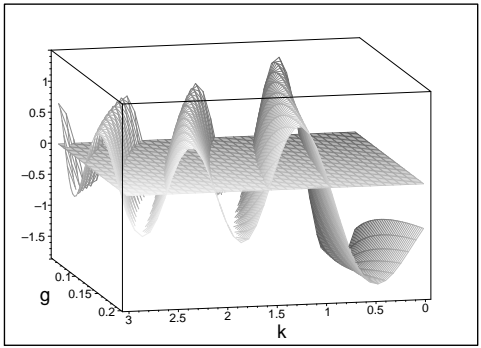,angle=270,width=0.36\textwidth}
\end{center}    
\vspace{2mm} \caption{The sample of the shape of the asymptotically
decreasing solution (\ref{hlejost}) at the small couplings $g$ (its
intersections with the horizontal plane define the physical energies
$E_n=k^2_n(g)$ as functions of $g$ at $n=1,3,5,\ldots$).
 \label{meto}
 }
\end{figure}

Via an explicit numerical test paying attention just to the simplest
secular Eq.~(\ref{oddsol}) in the odd-parity case, a definite
encouragement of our present optimism was provided by the picture
showing the smooth parameter-dependence of the asymptotically
decreasing solutions (see Fig.~\ref{meto}). Moreover, the
real-function nature of the picture illustrates that there exists a
suitable {\it ad hoc} normalization which keeps our asymptotically
decreasing function of variables $g$ and $k$ real.

%
\begin{figure}[h]                    
\begin{center}                         
\epsfig{file=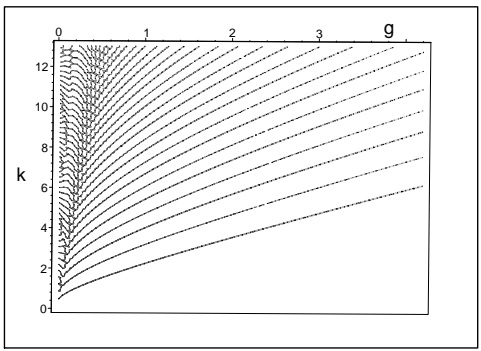,angle=270,width=0.36\textwidth}
\end{center}    
\vspace{2mm} \caption{The numerically determined odd-parity zeros
$k=k_n(g)$, $n=1,3,5,\ldots$ of the
asymptotically-decreasing-solution secular equation (\ref{oddsol}).
 \label{hladi}
 }
\end{figure}

The latter two observations (plus their parallels in the even-parity
case) explain the straightforward nature of the algorithms of the
numerical search for the bound-state roots of the exponential-well
secular equations. For this reason we were able to extend the
computations based on the asymptotically-decreasing-solution
algorithms beyond the small-coupling domain. The results are sampled
in Fig.~\ref{hladi}.

\subsection{Regular solutions and the both-sided estimates of the energies}

A more detailed inspection of the numerical results as presented in
Fig.~\ref{hladi} reveals that the use of the fixed-precision
arithmetics leads to the loss of the reliability of the localization
of the bound-state roots $k_n(g)$ in the domain of small
$g<g_{critical}(n)$. Empirically we see a more or less linear
$n-$dependence of $g_{critical}(n)$, with maximal $g_{critical}(45)
\approx 0.57$ at the highest level as identified in
Fig.~\ref{hladi}.

Naturally, such a loss of precision is not too surprising. In fact,
our {\it a priori} decision of preferring the asymptotically
decreasing solutions (\ref{hlejost}) proved rather lucky because
Fig.~\ref{hladi} appeared to provide a reliable information about
the spectrum in the more difficult and non-perturbative
``strong-coupling" dynamical domain.

This being said, the exact solvability of our radial-like
Schr\"{o}dinger equation still admits the use of an alternative
strategy in which one would {\em start} from the initial conditions
(\ref{evensec}) and (\ref{oddsec}) and in which one would {\em
construct} the so called regular solutions $\psi_{(regular)}(r,k^2)
$ by the purely analytic and non-numerical means. In such a setting,
naturally, the physical values of $k=\sqrt{E_n}$ will have to be
sought via the {\em fit} of the regular solutions to the standard
(i.e., Dirichlet) asymptotic boundary conditions,
 \be
 \psi_{(regular)}(R,k^2)=0\,,\ \ \ \ R \gg 1\,.
 \label{asyfit}
 \ee
One can expect that such an alternative construction strategy could
cover not only the domains of parameters in which the
above-described asymptotically-decreasing-solution approach had
failed but also some of the applications of the solutions in which
one needs a very precise evaluation of the wave functions.

Although the regular-solution
algorithm could start, in principle, from the same change of
variables as above, our analysis showed that the analytic
formulae putting emphasis on the matching at $r=0$ become simpler if
we change some signs and use the abbreviations
 \be
 K_\nu(z)=\frac{i\pi}{2}\exp \left (\frac{i\pi}{2}\nu\right
 )H^{(1)}_\nu(iz)
 \ee
 \be
 I_\nu(z)=\exp \left (-\frac{i\pi}{2}\nu\right
 )J_\nu\left (\exp \left (\frac{i\pi}{2}\right
 )z\right )\,.
 \ee
In this manner, the modified,  small$-r-$friendly ansatz
 \be
 \psi_{(regular)}(r,k^2)=
 D_1(\nu,g)\,K_{\nu}(w)+
 D_2(\nu,g)\,I_{\nu}(w)\,,
 \ \ \ \ w = w(g,r)=2ge^{r/2}\,,
 \ \ \ \ \nu = 2ik\,
 \label{ika}
 \ee
leads to the parity-dependent final results again. Thus, up to a
modifiable overall normalization the choice of the even parity may
be shown to yield, after the straightforward though still rather
tedious computations, the coefficients given by the
derivative-related formulae
 \be
 D_1(\nu,g)=g\,I_{\nu+1}(2g)+ik\,I_\nu(2g)\,,
 \ \ \ \
 D_2(\nu,g)=g\,K_{\nu+1}(2g)-ik\,K_\nu(2g)
 \,\ \ \ {\rm (even \ parity)}
 \,.
 \label{ikaplus}
 \ee
The choice of the odd parity leads to simpler expressions,
 \be
 D_1(\nu,g)=-I_\nu(2g)\,,
 \ \ \ \
 D_2(\nu,g)=K_\nu(2g)
 \,\ \ \ {\rm (odd - parity\ case)}
 \,.
 \label{ikaminus}
 \ee
Having these formulae at our disposal we are now prepared to test
the numerical performance of the regular-solution-based recipe.

At any suitable trial energy variable $E=k^2$ the recipe requires
that we formulate the secular equation for the physical bound-state
energies as the Dirichlet asymptotic boundary condition. In the
numerical practice this means that we preselect a suitable,
sufficiently large value $R$ of the coordinate and, in both the
even- and odd-parity cases, search for all of the roots $k=k_{n}(R)$
of transcendental Eq.~(\ref{asyfit}).
In the limit of large $R$ this form of secular equation should again
determine the exact physical values of the energy levels $E=E_n =
k_n^2(\infty)$.

%
\begin{figure}[h]                    
\begin{center}                         
\epsfig{file=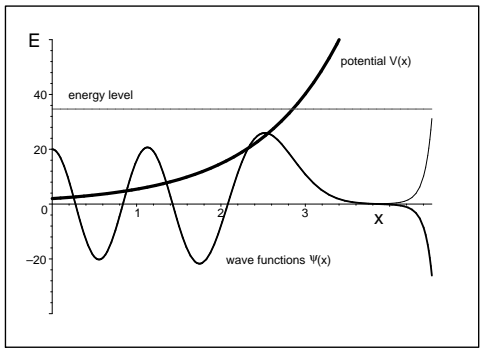,angle=270,width=0.36\textwidth}
\end{center}    
\vspace{2mm} \caption{The sample of $V(x)$ (at $g^2=2$), of one of
its highly-excited even-parity bound-state energy (viz., of $E_n$
with $n=8$) and of the related upper- and lower-bound radial-like
wave-function $\psi_n(r)$, with normalization at $r=0$ being
re-scaled here {\it ad hoc}.
 \label{edi}
 }
\end{figure}

As we already indicated the key appeal of the use of regular
solutions (\ref{ika}) lies, {\it a priori}, in a more friendly
representation of the wave functions. This expectation is confirmed
by Fig.~\ref{edi} in which the visible difference between the wave
functions evaluated at the minorizing and majorizing
$k=k_{physical}\pm 0.0001$ only becomes detectable beyond $r \approx
4$. In parallel, Table \ref{pexp4} shows that the sufficient cut-off
$R$ need not lie too far beyond the intersection $x_0$ of the energy
level with the exponentially growing wall of the potential well.

\begin{table}[h]
\caption{The sample of parameters for the numerical bracketing of
the even-parity energies using secular Eq.~(\ref{asyfit}) ($\psi(R)=
0$, $E=V(x_0)$). } \label{pexp4}

\vspace{2mm}

\centering
\begin{tabular}{||c||c|c|c|c||}
\hline \hline
    {\rm n} & $E_n^{(lower\ bound)}$ &
     $E_n^{(upper\ bound)}$&  $R$ & $x_0$
    \\
    \hline \hline
 0&  4.12005  & 4.12010& 3.0 & 0.72\\
 2&  11.0065  & 11.0075& 3.0 & 1.71\\
 4&  18.2822  & 18.2830& 3.4 & 2.21\\
 \hline
 \hline
\end{tabular}
\end{table}

\section{Discussion\label{discussion}}

In the present note we demonstrated that the elementary transition
from the full-line equation (\ref{SEx}) to the half-line equation
(\ref{SExexright}) enables one to claim that in spite of the
admitted non-analyticity in the origin our present model can be
perceived as exactly solvable. In other words, we believe that the
conventional families of the exactly solvable one-dimensional
potentials (with their typical list provided by review paper
\cite{Cooper}) should be complemented by the symmetric functions
$V(x)=V(-x)$ which are non-analytic in the origin but still
tractable via special functions (cf. also several other comments
\cite{rese,Souza,Ishkh,Quesne,Quesneb,Quesnec,Quesned} in this
respect).


Once more, let us remind the readers that our choice and study of
potential (\ref{exwe}) was motivated by the need of interpolation
between the ubiquitous harmonic oscillator $V^{(HO)}(x)=
\omega^2x^2$ and its equally easily solvable square-well alternative
 \ben
 V^{(SW)}(x)=
 \left \{
 \begin{array}{llc}
 \infty\,,\ \ & x < -L,\\
 0\,,          & x \in (-L,L)\,,\\
 \infty\,,&x>L
 \ea
 \right .
 \een
with, say, $L = 1/\omega$. Such a need of interpolation may have
various pragmatic as well as theoretical reasons \cite{Uwe}.
Nevertheless, one of the main difficulties encountered during such a
search is usually seen in an inadvertent loss of the appealing,
textbook-explained exact solvability (ES) status of the underlying
one-dimensional bound-state Schr\"{o}dinger Eq.~(\ref{SEx}).

In practice, due to the ordinary linear differential nature of
Eq.~(\ref{SEx}) it is still fortunate that one can employ standard
numerical methods and one can construct the bound state solutions,
with arbitrarily predetermined precision, for a vast majority of the
potentials of phenomenological interest. Still, one usually argues
that the strictly non-numerical nature of bound states makes the ES
interactions $V^{(HO)}(x)$ and $V^{(SW)}(x)$, in many a respect,
privileged.

By our opinion, the key weakness of the latter argument lies in the
vagueness of the very definition of the ES status. This definition
proves even different for the smooth, analytic potentials and for
the various discontinuous versions and descendants of square wells
for which the wave functions remain elementary but for which, in a
way illustrated, e.g., in Refs.~\cite{sqw,sqwb,sqwc,sqwd}, the
energies must still be sought -- purely numerically -- as roots of
transcendental equations. In the former case, on the contrary, the
energies are usually given by closed and elementary formulae while
the wave functions themselves remain elementary only due to a
slightly mysterious degeneracy of the infinite-Taylor-series
special-function general solutions to classical orthogonal
polynomials (cf., e.g., \cite{Cooper} for a representative list of
the analytic ES interaction models as sampled here by
$V^{(HO)}(x)$).

As a true test of the robustness of the ES concept we studied here
the ``non-ES'' toy-model interaction (\ref{exwe}) which can be also
considered, {\it cum grano salis}, solvable. In addition, potential
(\ref{exwe}) shares certain geometrical as well as solvability
features with {\em both} of the above-mentioned ES examples. At the
small couplings $g^2>0$ and in the low-lying energy region its shape
resembles harmonic oscillator (cf. Fig.~\ref{aqizt} where
$g^2=1/4$). With the growth of the coupling the spike at $x=0$
becomes more pronounced (return to Fig.~\ref{reqizt} where the value
of $g^2$ varied from 1/4 to 4).

It is obvious that with the growth of $g^2$ the spectrum is being
pushed upwards. Moreover, at higher excitations the function
(\ref{exwe}) of $x$ becomes steeper so that the potential resembles,
more and more, square well. In contrast to the equidistant
harmonic-oscillator case the $n \gg 1$ differences
$\delta_n=E_{n+1}-E_n$ between the neighboring levels may be,
therefore, expected to grow - more or less quadratically - with $n$.
Indeed, such an expectation is numerically confirmed by
Fig.~\ref{udine}

%
\begin{figure}[h]                    
\begin{center}                         
\epsfig{file=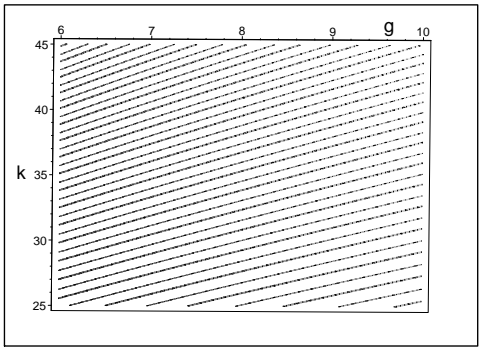,angle=270,width=0.36\textwidth}
\end{center}    
\vspace{2mm} \caption{The sample of the approximate equidistance of
the square roots $k_n=\sqrt{E_n}$ of the highly-excited even-parity
bound-state energies in the domain of $g \in (6,10)$ and $k \in
(25,40)$.
 \label{udine}
 }
\end{figure}

From the purely phenomenological point of view interaction
(\ref{exwe}) is made interesting by an interplay between the latter
two features. The model seems also remarkable on the purely formal
grounds because, as we showed, the underlying bound-state
Schr\"{o}dinger equation (\ref{SEx}) proves piecewise exactly
solvable in terms of the various forms of Bessel functions.
Naturally, this could open a way towards various further
generalizations.

Last but not least we would like to emphasize  that we also showed
here how one can circumvent the usual precision-fighting
difficulties. Incidentally, some of these difficulties are
intimately connected with the extremely quick exponential growth of
the walls of our potential well (\ref{exwe}) at large distances. In
this context we may also recall Ref.~\cite{Flueggedeut} in which, in
the context of a simplified description of the deuteron, a similar
necessity of the cross-checking verification of the numerical
reliability of the exact $s-$wave formulae has been noticed to occur
in the opposite extreme of the exponentially decreasing potentials.

\subsection*{Acknowledgements}

The project was supported by IRP RVO61389005 and by GA\v{C}R Grant
Nr. 16-22945S.

\end{document}